\documentclass[10pt, preprint2]{aastex}

\shorttitle{The Hubble constant inferred from 18 time-delay lenses}
\shortauthors{Paraficz et al.}

\begin{document}

\title{The Hubble constant inferred from 18 time-delay lenses}

\author{Danuta Paraficz  and Jens Hjorth}
\affil{Dark Cosmology Centre, Niels Bohr Institute, University of Copenhagen,
		Juliane Maries Vej 30, DK-2100 Copenhagen, Denmark}


\begin{abstract}
We present a simultaneous analysis of 18 galaxy lenses with time delay measurements. For each lens  we derive mass maps using pixelated simultaneous  modeling with shared Hubble constant. We estimate the Hubble constant to be $66_{-4}^{+6}$ $\rm{km}$ $\rm{s}^{-1} \rm{Mpc}^{-1}$ (for a flat Universe with $\Omega_m=0.3$, $\Omega_\Lambda=0.7$). 
 We have also selected a subsample of five  relatively isolated early type galaxies and by simultaneous modeling with an additional constraint on isothermality of their mass  profiles we get $H_0=76_{ -3}^{+3}$  $\rm{km}$ $\rm{s}^{-1} \rm{Mpc}^{-1}$.
\end{abstract}



\section{Introduction}
 
The Hubble constant is one of the most important parameters in 
 cosmology.  It determines the age of universe, the physical distances to objects, constrains the dark energy equation-of-state  and furthermore, it is used as a prior in many cosmological analysis. Hence, it is essential for cosmology to know the precise value of $H_0$  \citep{Riess:2009}. The newest measurements give a quite well defined $H_0$ with   estimated errors at the 5-10\% level but unfortunately $H_0$ differs among different cosmological methods and there is only marginal consistency between their $1 \sigma$ errors (see Table 1). 
 
Moreover, those methods are based on different physical principles and  more importantly they measure different consequences  of $H_0$. Both SNe and Cepheids measure luminosity distance but at different scales (distant and local Universe respectively). Cepheids can provide a luminosity distance  via the period-luminosity relation  but only in the local Universe, on the other hand, the significantly brighter SNe with their characteristic peak luminosity can measure cosmological distances but need to be calibrated with Cepheids. The SZ effect, which is based on high energy electrons  in a galaxy cluster distorting the CMB through inverse Compton scattering,  is proportional to the gas density of the galaxy cluster and that combined with the cluster's X-ray flux gives an estimate of the angular diameter distance. 
 Finally,  the angular power spectrum of the CMB gives information about many cosmological parameters which have correlated effects on the power spectrum, especially strong degeneracies are between $H_0$ and $\Omega_m$ and $\Omega_b$. Furthermore, there are significant differences in the obtained results even within one  method (eg. SNe \citet{Branch:1996, Sandage:2006, Riess:2009}, see Table 1).
 
Additionally,  any astrophysical approach suffers from systematic uncertainties, e.g.,
supernovae are not standard but standardized candles with possible redshift evolution, CMB have various parameter degeneracies and interference with foregrounds, and the SZ method is assuming spherical symmetry on often significantly non-spherical clusters of galaxies. It is therefore important to explore complementary methods for measuring $H_0$.  Gravitationally lensed quasars QSOs offer such an attractive alternative. 

\begin{table}
\caption{$H_0$ comparison}
\begin{tabular}{ccc}
\hline\hline
Method&$H_0$& Author\\
\hline
CMB (3 years)&$73.2^{+3.1}_{-3.2}$&\citet{Spergel:2007}\\
CMB (5 years)&$71.9^{+2.6}_{-2.7}$&\citet{Komatsu:2009}\\
Ia SNa&$74.2 \pm 3.6 $&\citet{Riess:2009}\\
Ia SNa& $62.3 \pm 1.3$&\citet{Sandage:2006}\\
SZ & $65^{+8}_{-7}$&\citet{Jones:2005}\\
Cepheids&$72\pm8$&\citet{Freedman:2001}\\
\hline
\end{tabular}
\end{table}

 As shown by \citet{Refsdal:1964} the Hubble constant can be measured based on the time delay $\Delta t$ between multiply lensed images of QSOs, because $H_0\propto 1/\Delta t$, provided that the mass distribution of the lens is known.  Time delays measure $\frac{D_{OL}D_{OS}}{D_{LS}}$, where $D_{\rm OL}$, $D_{\rm OS}$, $D_{\rm LS}$ are the angular diameter distances between observer and lens, observer and source, and lens and source, respectively. 
Gravitational lensing has its degeneracies but it is based on well understood physics and unlike distance ladder methods there are no calibration issues \citep{Branch:1996, Sandage:2006}.

Gravitational lensing has, up to now, not been  seen as reliable as other leading cosmological methods. Determination of the Hubble constant using lensing is problematic, because the mass distribution of a lens strongly  influences the result of  $H_0$ and unfortunately we never have a complete knowledge of that, hence a choice of a lens model is needed.

Recently, however,  time-delay lenses have successfully  been used for $H_0$ estimation. In particular, \citet{Oguri:2007} used a Monte Carlo method to combine lenses and derived $H_0=70\pm 6$ $\rm{km}$ $\rm{s}^{-1} \rm{Mpc}^{-1}$. \citet{Saha:2006} obtained $H_0=72^{+8}_{-11}$ $\rm{km}$ $\rm{s}^{-1} \rm{Mpc}^{-1}$ using a combination of 10 lenses,  \citet{Coles:2008} got $H_0=71^{+6}_{-8}$ $\rm{km}$ $\rm{s}^{-1} \rm{Mpc}^{-1}$ using a combination of 11 lenses,  and \citet{Suyu:2009} found $H_0=69.7^{+4.9}_{-5.0}$ $\rm{km}$ $\rm{s}^{-1} \rm{Mpc}^{-1}$ by detailed analysis of one gravitational lens, B1608+656.

 This paper extends the work of  \citet{Saha:2006} and \citet{Coles:2008}, where results on $H_0$  were presented using combined modeling of 10 and 11 lenses, respectively. We have used those systems with refined properties of lenses as a part of our sample and have added  new systems that have been discovered and monitored during the past 4 years. We therefore now posses an almost doubled sample of systems with measured time delay, which demonstrates that, gravitational lensing is valuable method for $H_0$ estimation.


\section{Pixelated modeling}
Two different approaches for modeling lenses are commonly used.
The first one, the non-parametric (pixelated) method,  generates a large number of models which perfectly fit the data, each of them giving a different result  which can then be averaged. 
For the second method, model fitting, one assumes parametrized models of the mass distribution of the lens.

Pixelated modeling has the advantage of allowing the lens shape and profile to vary freely. It does not presume any parameters and
can provide models that would not be possible to reproduce with parametric modeling.
Using this approach to a combined analysis of a large sample of lenses is a powerful solution to the modeling problem in gravitational lensing.

In this paper we use the non-analytical method created by \citet{Saha:2004} - PixeLens.
PixeLens generates an ensemble of lens models that fit the input data. Each model consists of a set of  discrete mass points, the position of  the source   and optionally, if the time delays are known,  $H_0$. 
The time delay $ \Delta\tau$ is the combined effect of  the difference in length of the
optical path between two images and the gravitational time dilation of two light rays passing through different parts of the lens potential well,
 \begin{equation}
  \Delta\tau = \frac{1+z_{\rm L}}{c}\frac{D_{\rm OS}D_{\rm OL}}{D_{\rm LS}}
  \left( \frac{1}{2} (\vec{\theta}-\vec{\beta})^2-\Psi(\vec{\theta})\right).
  \end{equation}
Here $\vec{\theta}$ and $\vec{\beta}$ are the positions of the images and the source respectively, $z_{\rm L}$ is the lens redshift, and $\Psi$ is the effective gravitational potential of the lens.

The arrival time at position $\vec{\theta}$ is defined in PixeLens, as a modeled surface,
 \begin{equation}
 \tau(\vec{\theta}) = \frac{1}{2} |\vec{\theta}|^2-\vec{\theta} \cdot \vec{\beta}- \int\ln|\vec{\theta}-\vec{\theta}'|\kappa(\vec{\theta}')d^2\vec{\theta}',
   \end{equation}
   where $\kappa$ is surface mass density.

The errors of the positions of observed images and redshifts of source and
lens are of the order of a few percent, thus can be ignored. The main source of errors in the data comes from time delays between images.

We have used PixeLens  to generate a set of 100 models for a sample of  lensing systems (see \S 3).
 PixeLens produces an ensemble of models with varying $H_0$, each consisting of sets of mass pixels,   which exactly reproduce  the input data. Moreover, it also models several lenses simultaneously enforcing shared $H_0$ for all lenses.
  The images positions, the source and lens redshifts, and the time delay are assumed to be accurate enough for their errors  to be ignored \citep{Saha:2006}.
PixeLens also imposes secondary constraints on mass maps: 
non-negative density; smoothness, where the density of a pixel must be no more than twice the average density of its neighbors; the mass profile is required to have $180^\circ$ rotation symmetry (except if it  appears very asymmetric); the shear is allowed within $45^\circ$ of the chosen direction;  circularly averaged mass profile should nowhere be  shallower nor steeper than $r^{-\alpha_{min}}$ and $r^{-\alpha_{max}}$ respectively, where $\alpha_{min}$ and  $\alpha_{max}$ is defined in PixeLens as the minimum and maximum steepness, those values can be chosen by the user however, the default PixeLens constraint is minimal steepness $\alpha_{min}=0.5$;  and finally, additional lenses as point masses can be constrained.
 PixeLens does not use  flux ratios as constraints because of the possible influence of reddening by dust \citep{Eliasdottir:2006}, microlensing \citep{Paraficz:2006} or small-scale structure in the lens potential \citep{Dalal:2002}.

\section{Data set}

To date, there are 19 gravitational lens systems with published time delays. 
Table 2 summarizes  the information about these 19 systems.
We have made an attempt to use all the conclusions previously drawn about their shape, external shear, profile, etc.
We have used the newest/best measurement of positions and redshifts of images and lens. Apart from the main lensing galaxies we have also included all the galaxies that might contribute to the lensing. We added them whenever they are visible in the field. These systems are RX J0911+055, HE 1104-181, SBS 1520+530,  B1600+434 and B1608+656. 
All the mass maps of the doubly imaged quasars are required to have $180^\circ$ rotation symmetry and in case of the quadruply imaged systems we allow the lens to be asymmetric if it has been reported asymmetric,  which is the case for: HE 0435-1223, SDSS J1004+4112,  RX J1131-1231, B1608+656.
A constant external shear is allowed, for the lenses where the morphology shows evidence of external shear or the existence of  external shear has been reported, which is the case for HE 0435-1223, RX J0911+055,  FBQ J0957+561, SDSS J1004+4112, PG 1115+080,  RX J1131-1231, SDSS J1206+4332, B1600+434, SDSS J1650+4251, WFI J2033-4723.

One lens systems has been excluded from our analysis, B1422+231.
\citet{Raychaudhury:2003} indicated that the time-delay measurements made by \citet{Patnaik:2001} are possibly  inaccurate.
 \citet{Patnaik:2001} reported $\Delta T_{12} = 7.6 \pm2.5$ days, whereas \citet{Raychaudhury:2003} lens modeling predicts
$\Delta T_{12}=0.4 h^{-1}$ days.  According to \citet{Raychaudhury:2003} 
this value
would not be expected to show up in the \citet{Patnaik:2001} data, which sampled every 4 days. We follow the \citet{Raychaudhury:2003} prediction also because in our analysis 
the system gives an unreasonably low Hubble constant $H_0=12 \pm3$ $\rm{km}$ $\rm{s}^{-1} \rm{Mpc}^{-1}$, and we exclude the system in what follows.

  Figure 1 shows a mosaic of the average mass distributions for the remaining 18 lenses. 
  
\section{Full set results}

Our resulting $H_0$ distribution is shown in Figure \ref{fig3}. We have performed the calculation of 18 systems for a flat  Universe with $\Omega_\Lambda=0.7$, $\Omega_m=0.3$ and we obtain $H_0=66_{-4}^{+6}$  $\rm{km}$ $\rm{s}^{-1} \rm{Mpc}^{-1}$ at 68\% confidence and  $H_0=66_{-7}^{+8}$  $\rm{km}$ $\rm{s}^{-1} \rm{Mpc}^{-1}$ at 90\% confidence.
We also note that for $\langle z_L \rangle=0.6$, $\langle z_S \rangle=1.8$, that are average lens and source redshifts of our sample, the inferred $H_0$ should increase by 2\% for an open Universe with $\Omega_\Lambda=0.0$, $\Omega_m=0.3$ and decrease by 7\% for an Einstein-de Sitter Universe $\Omega_\Lambda=0.0$, $\Omega_m=1.0$.

Figure 3 presents the comparison between estimation from our sample of lenses and samples from previous lensing studies using 15 lenses\footnote{Excluding  B1422+231.} 
\begin{figure*}[!h]
\epsscale{.40}
\plotone{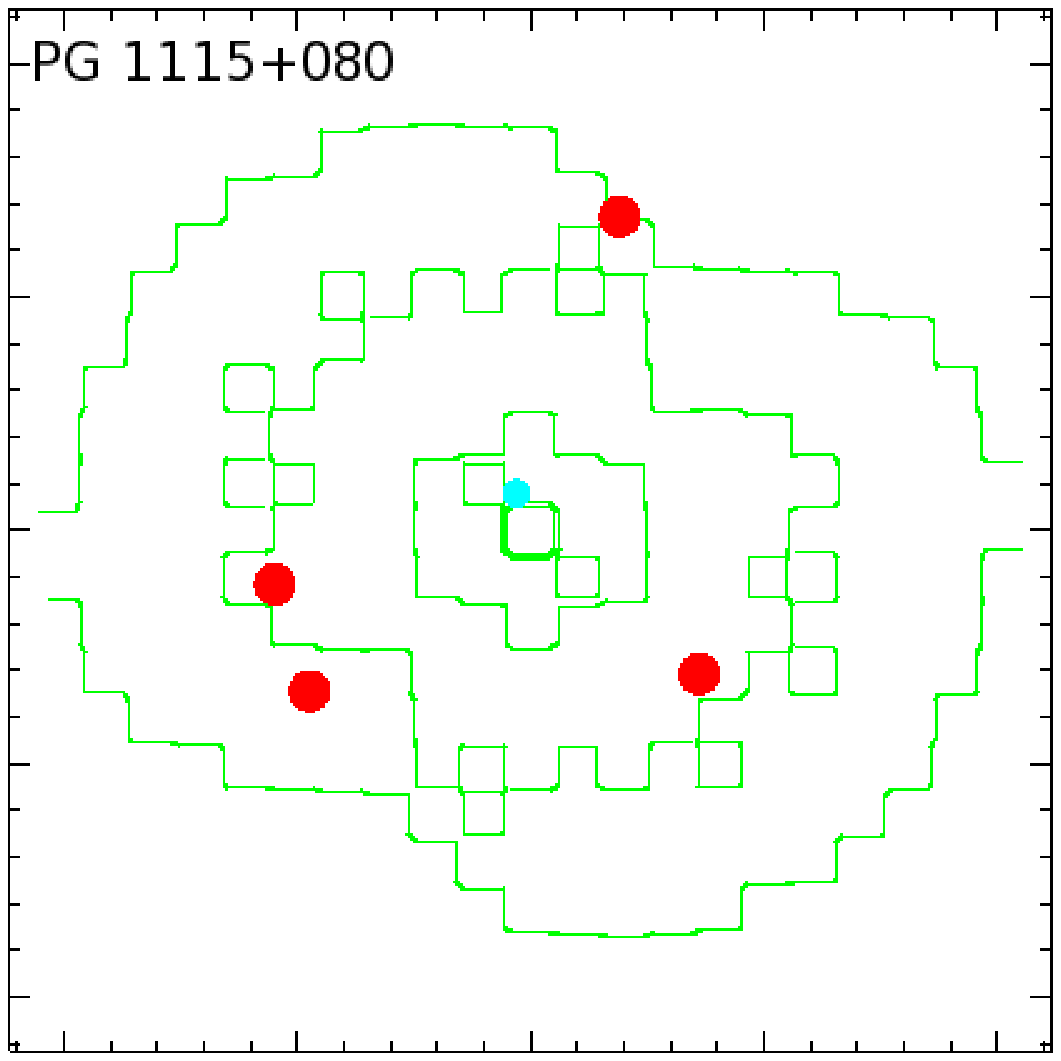}\plotone{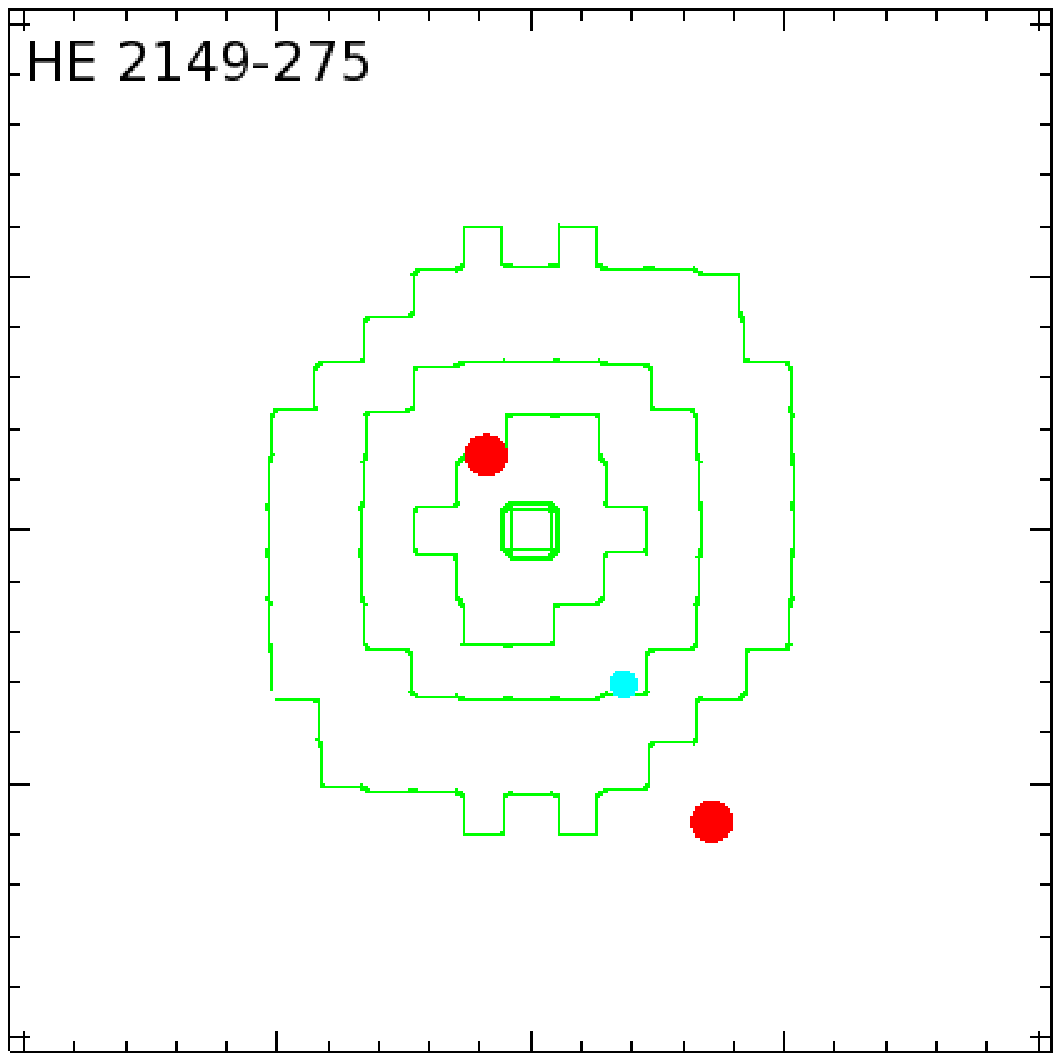}\plotone{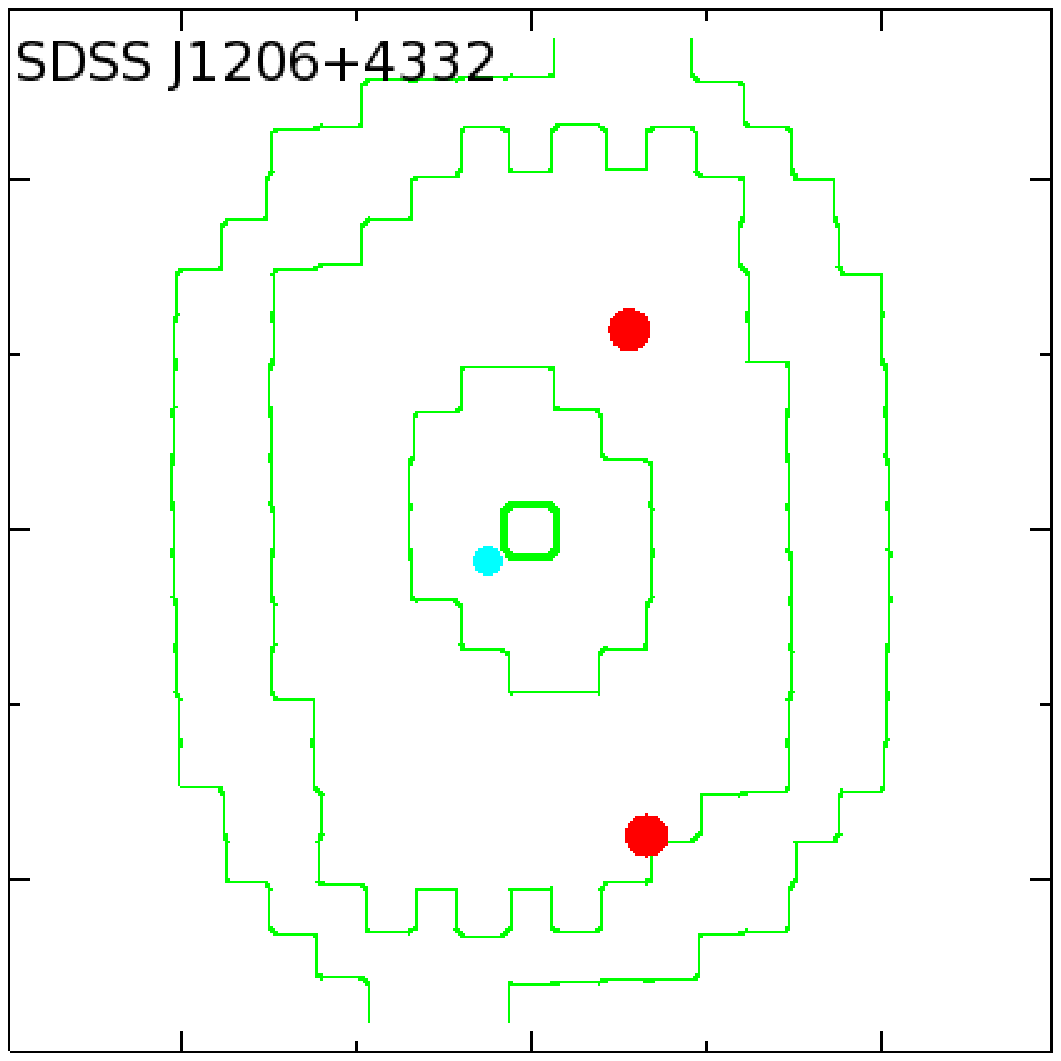}\\
\plotone{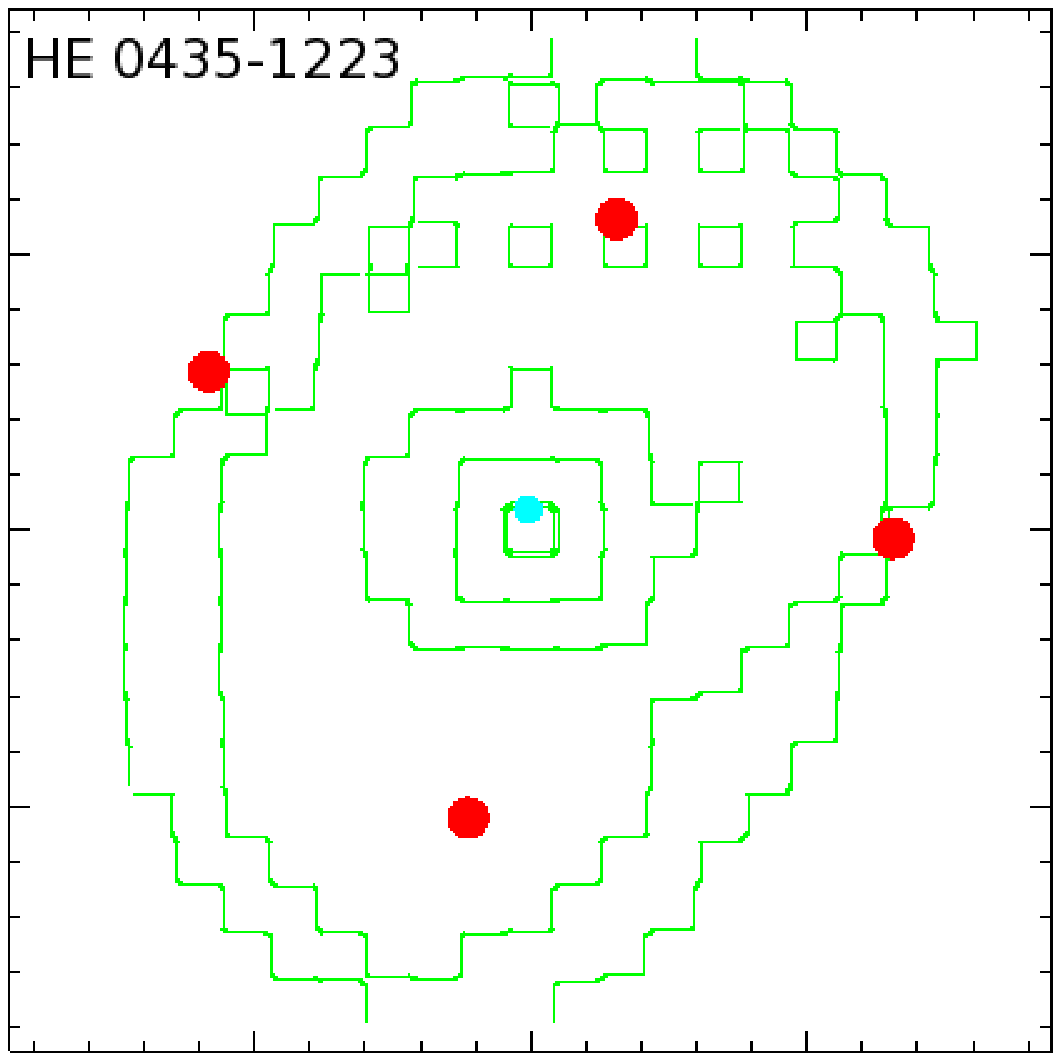}\plotone{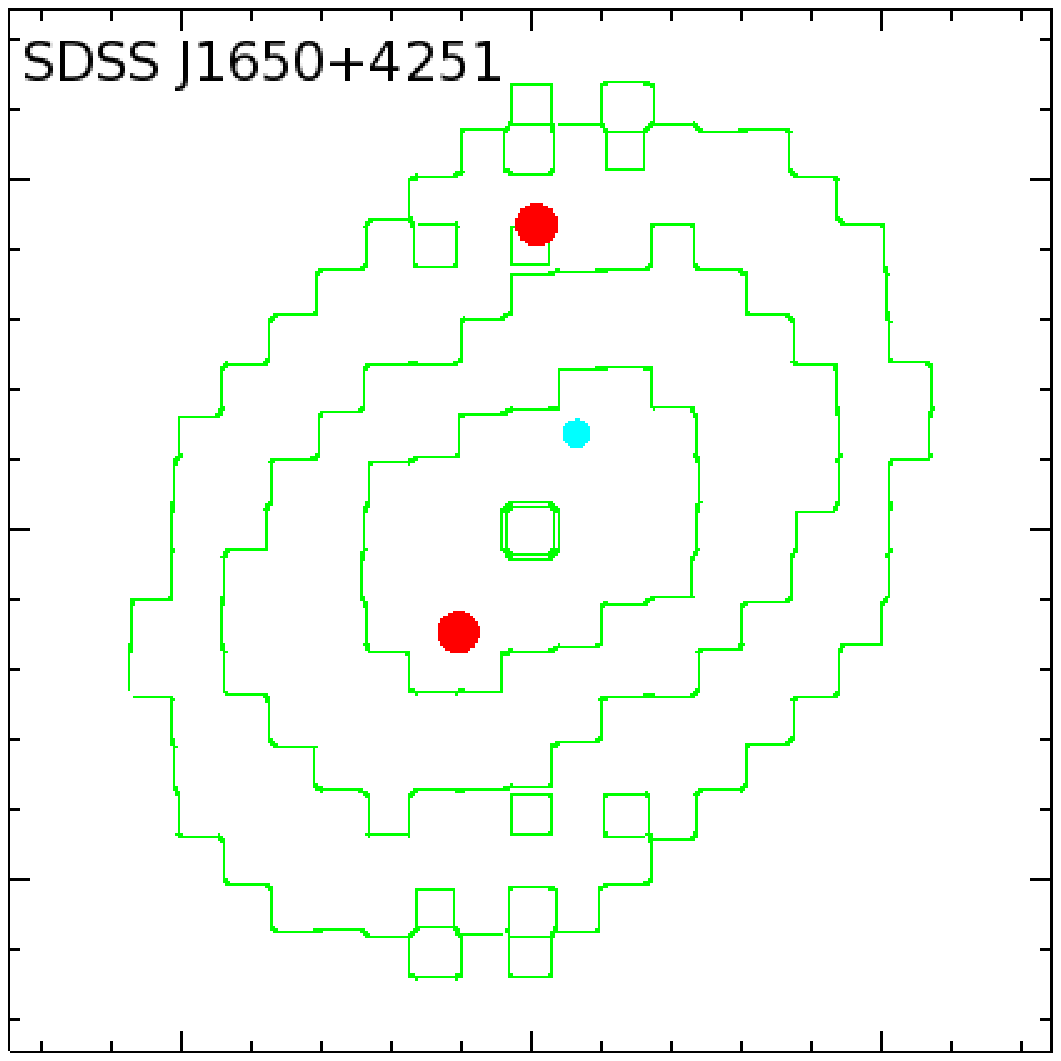}\plotone{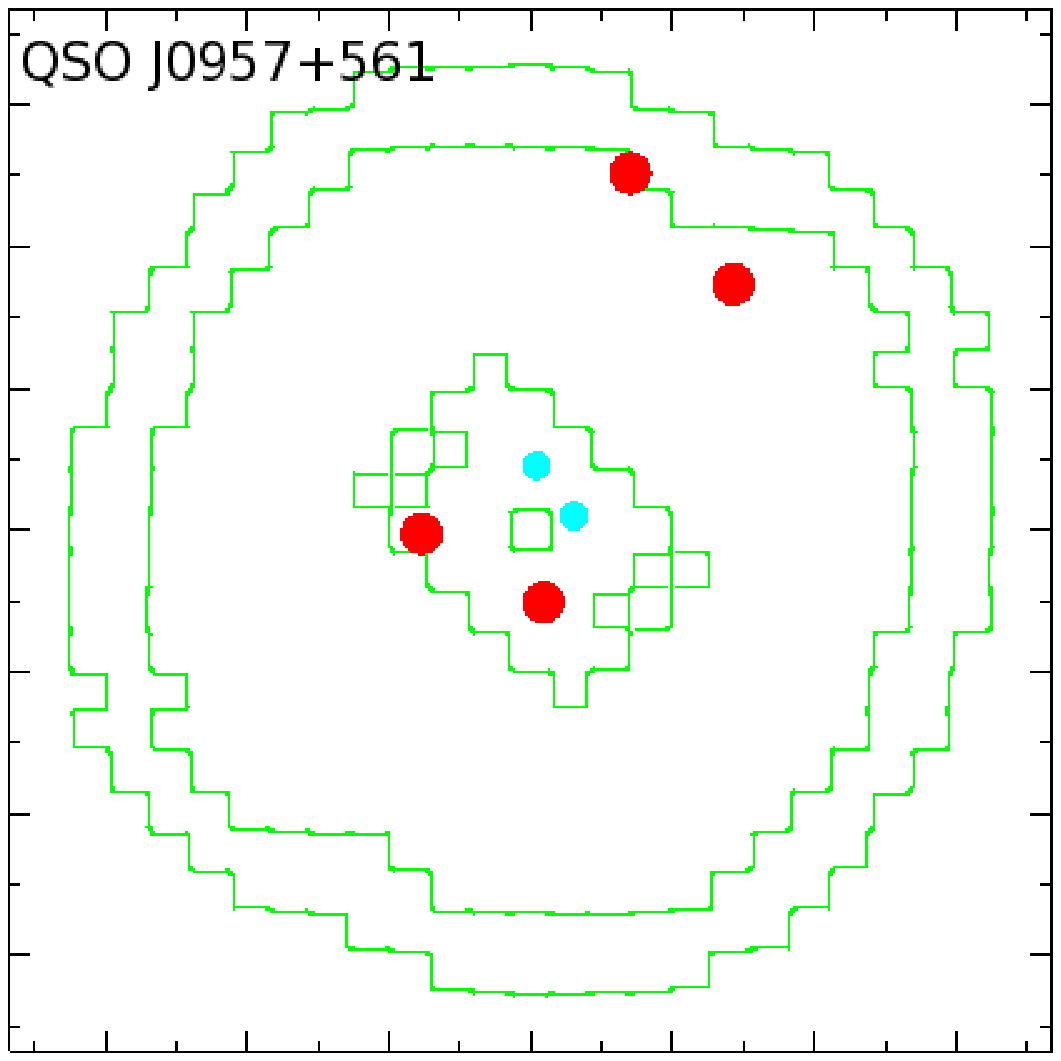}\\
\plotone{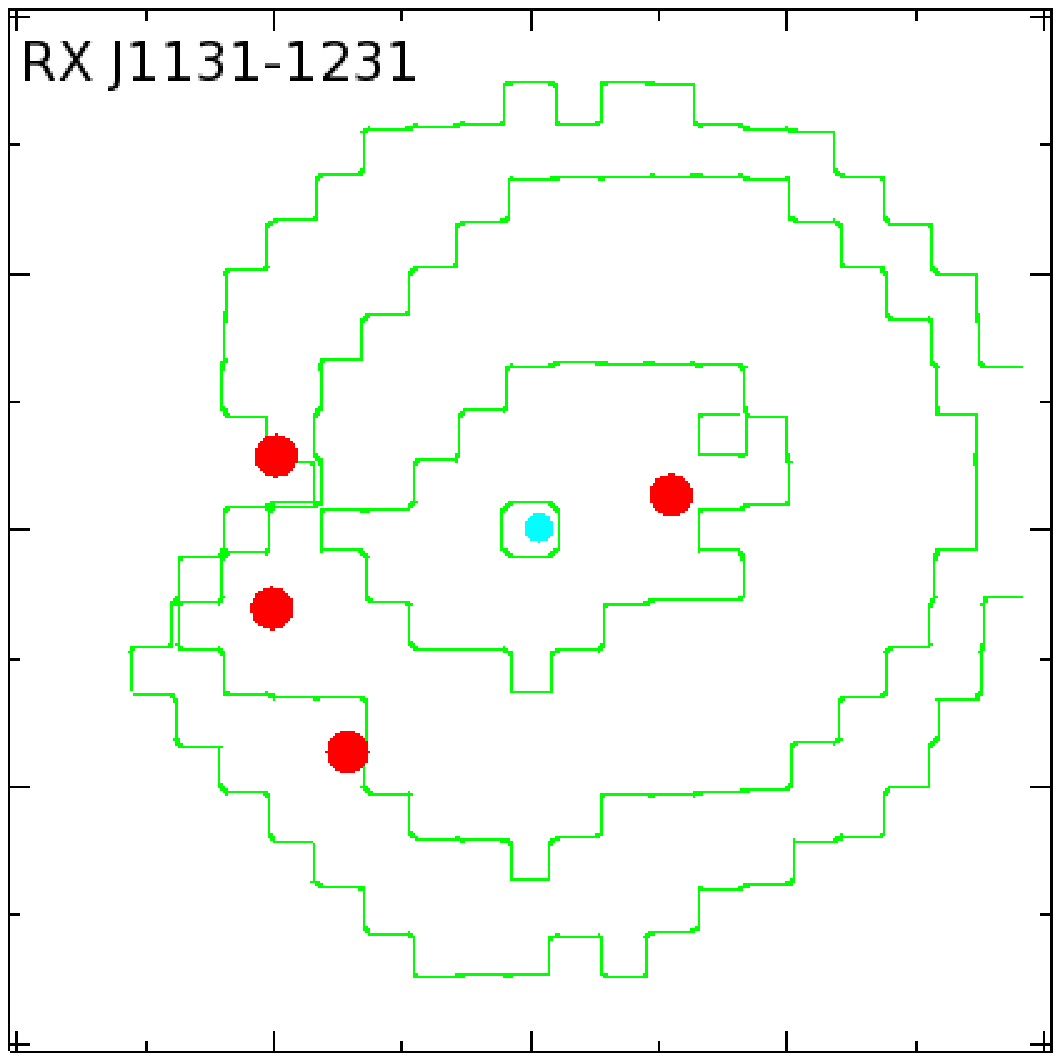}\plotone{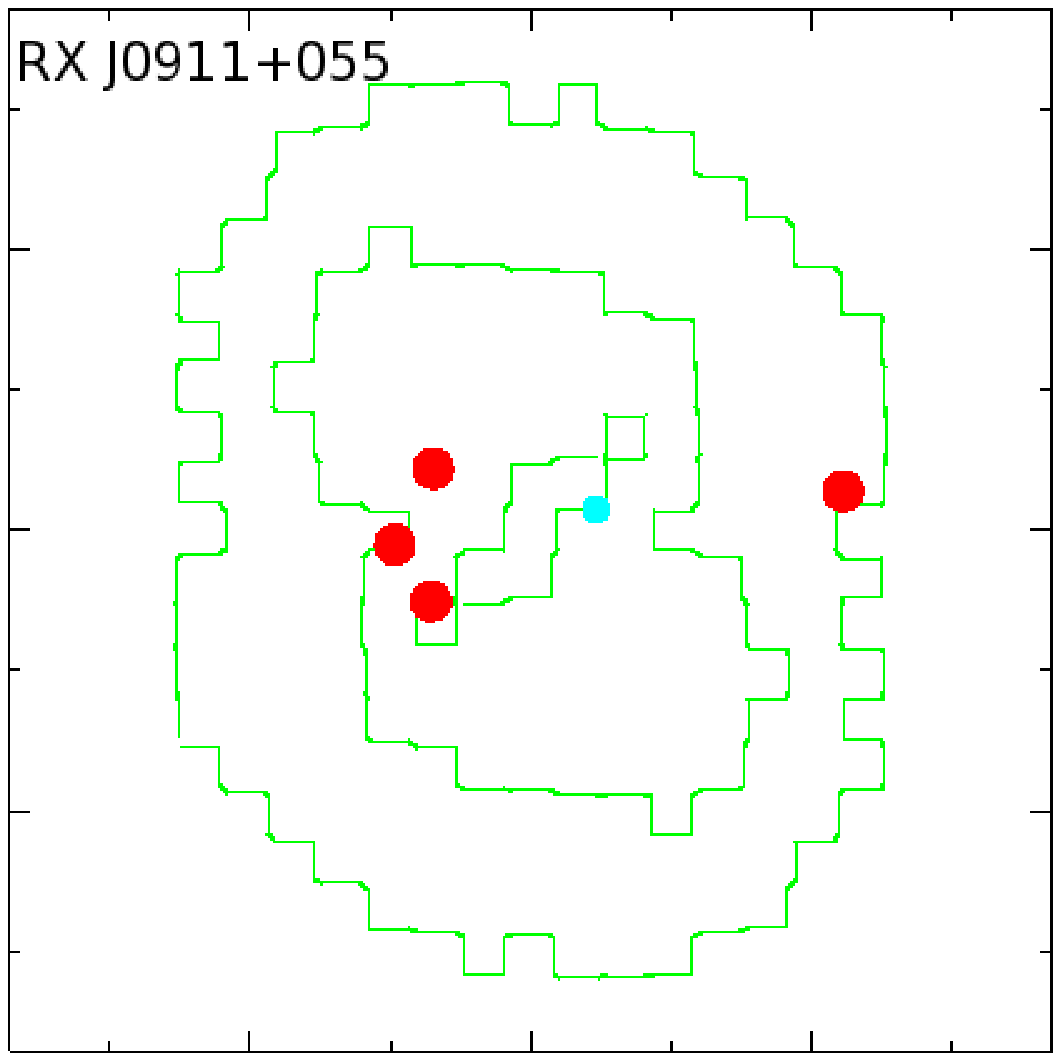}\plotone{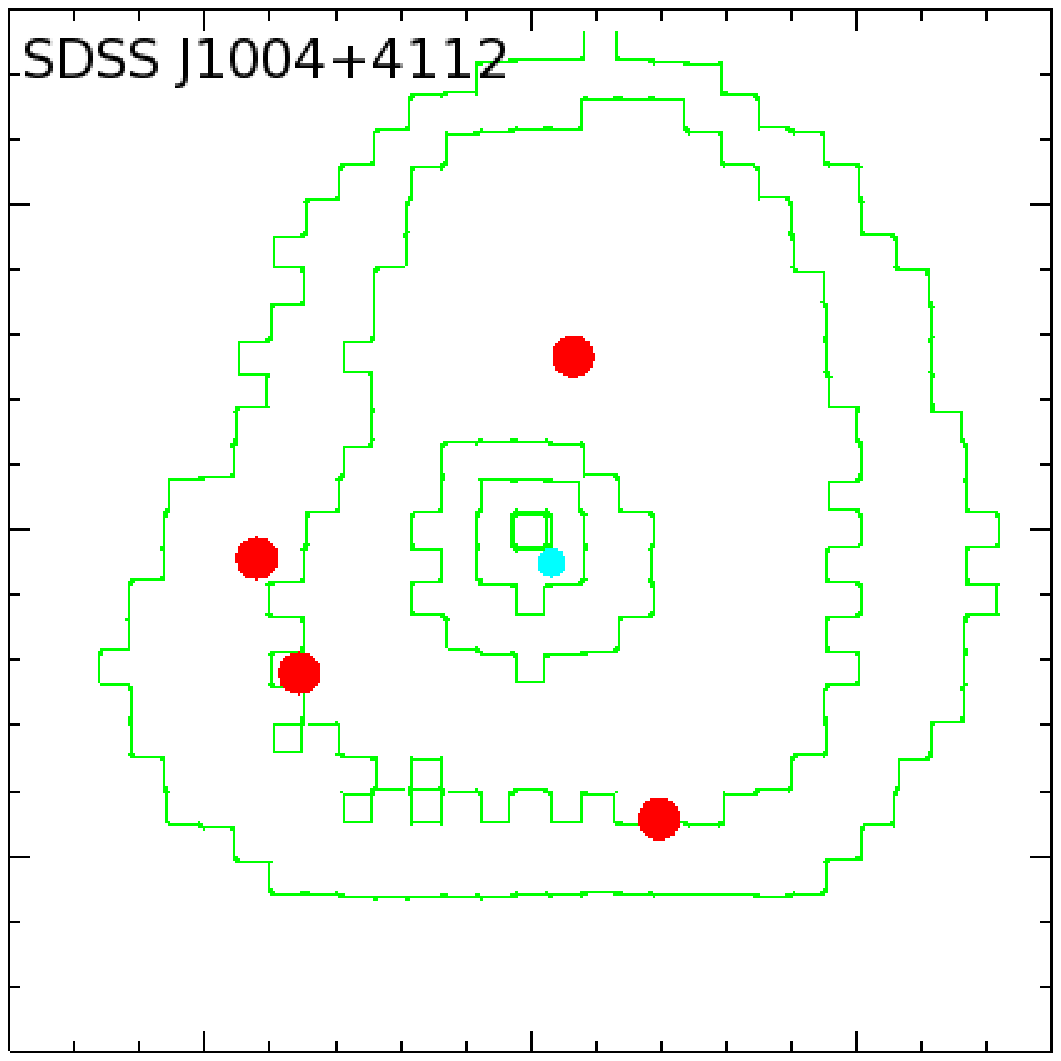}\\
\plotone{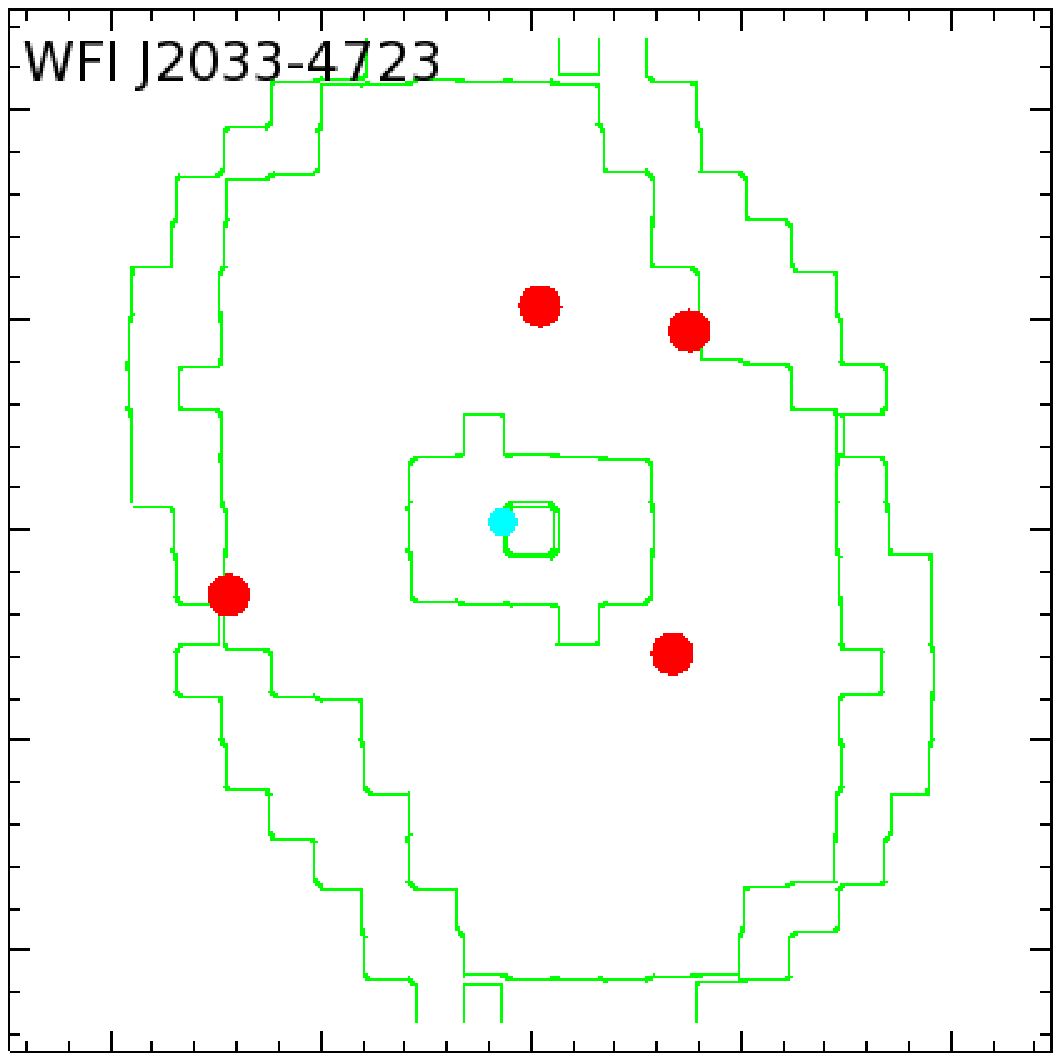}\plotone{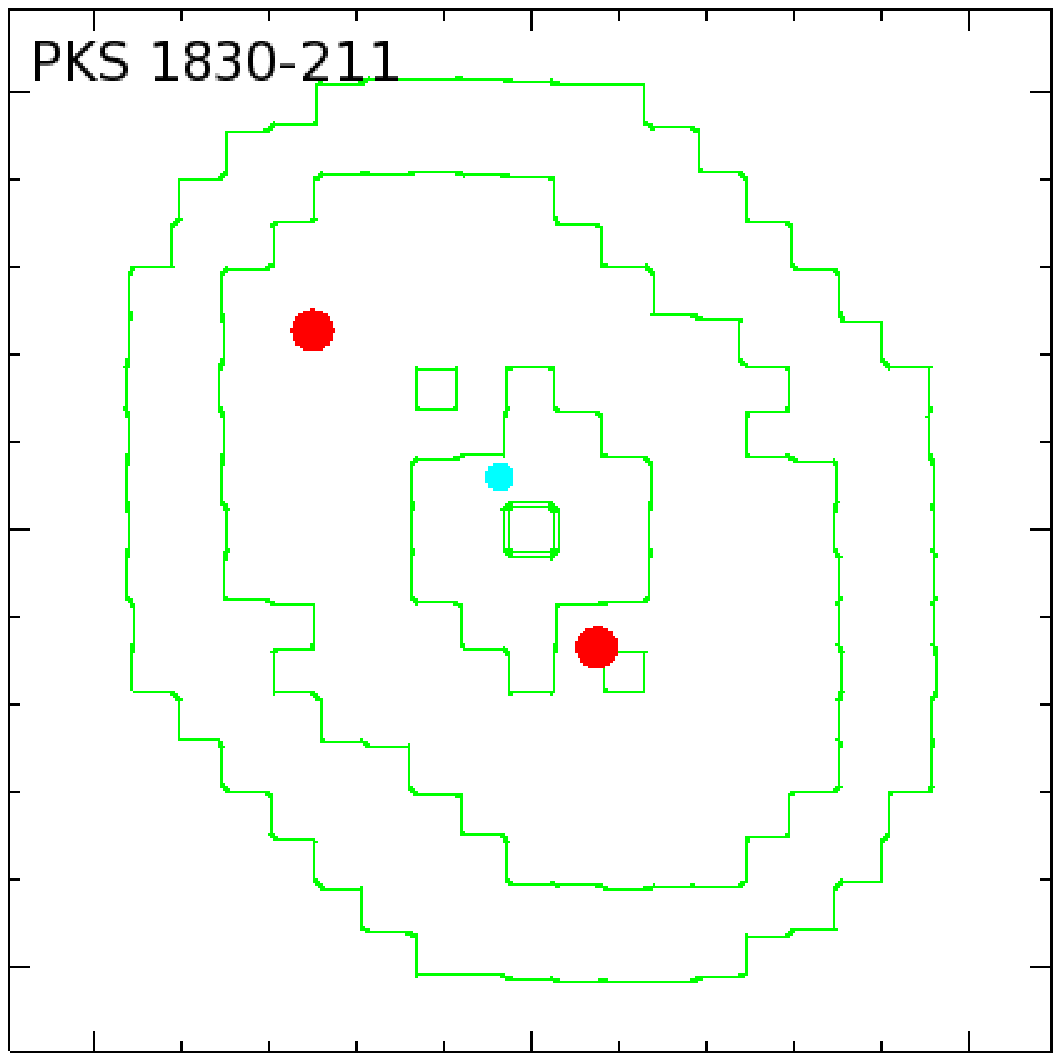}\plotone{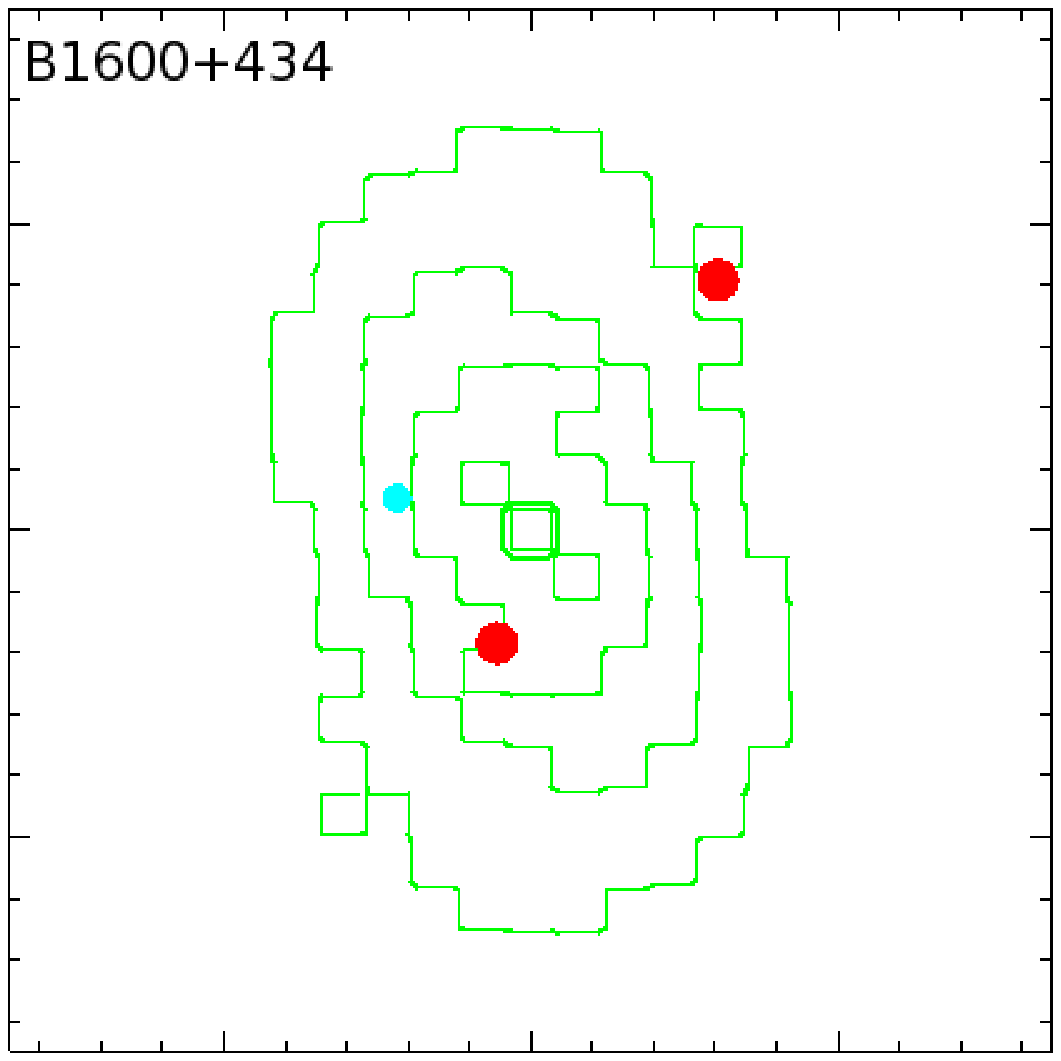}\\
\plotone{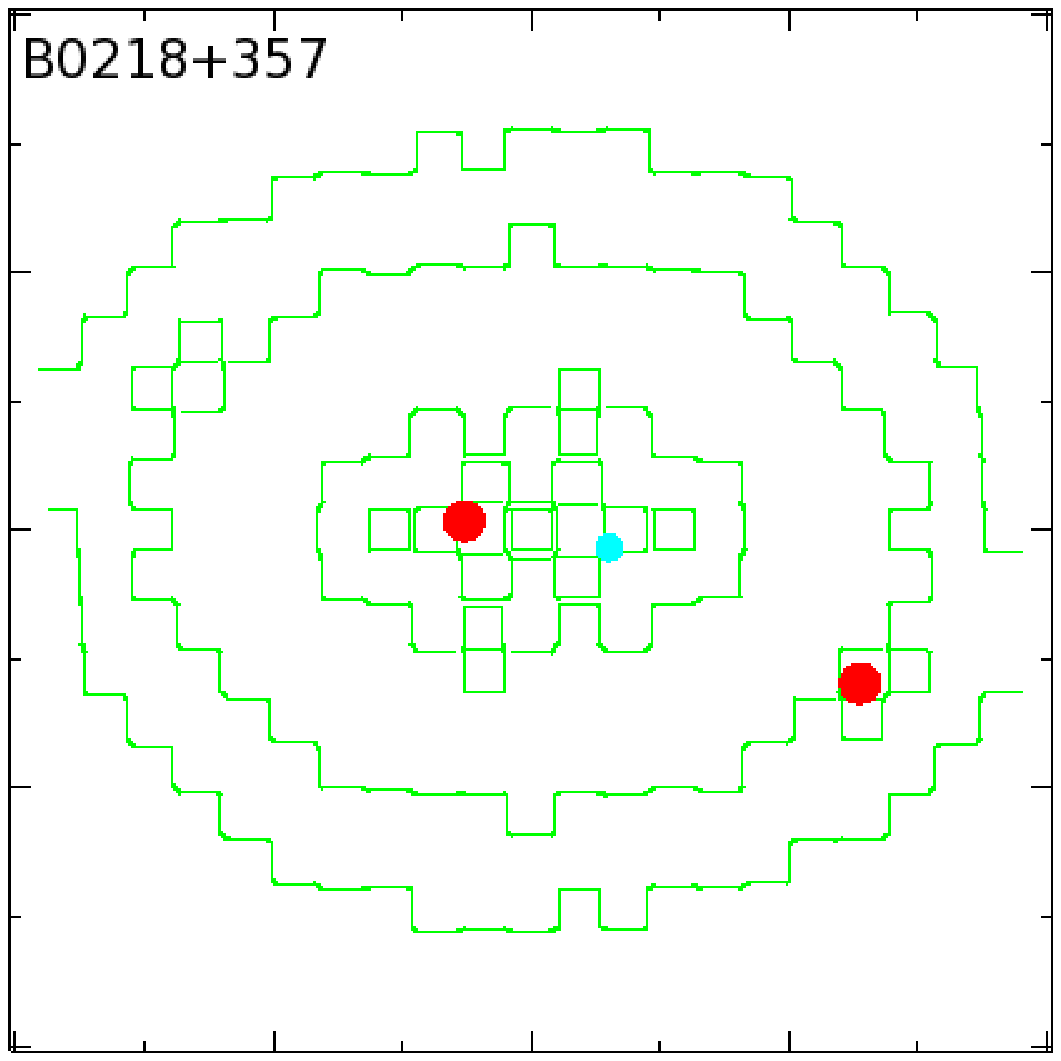}\plotone{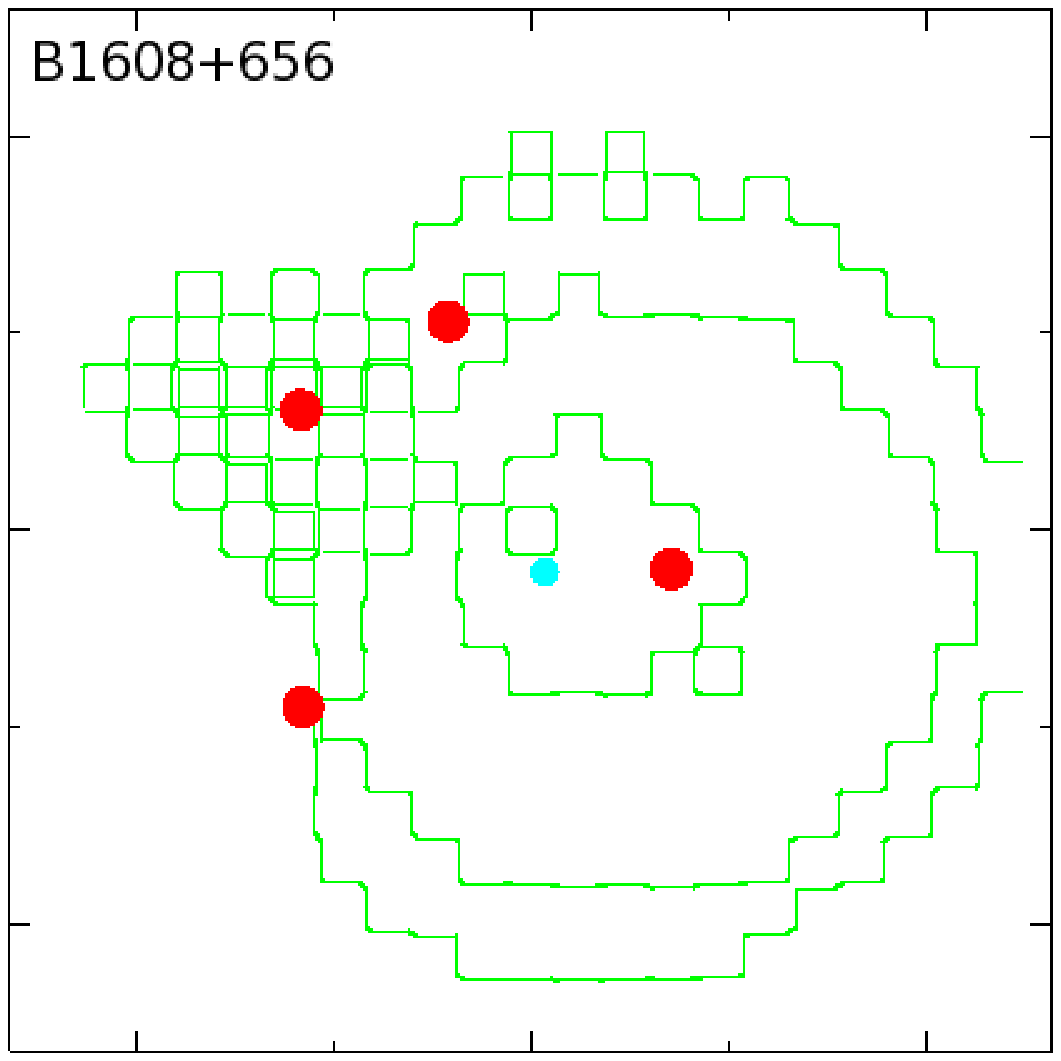}\plotone{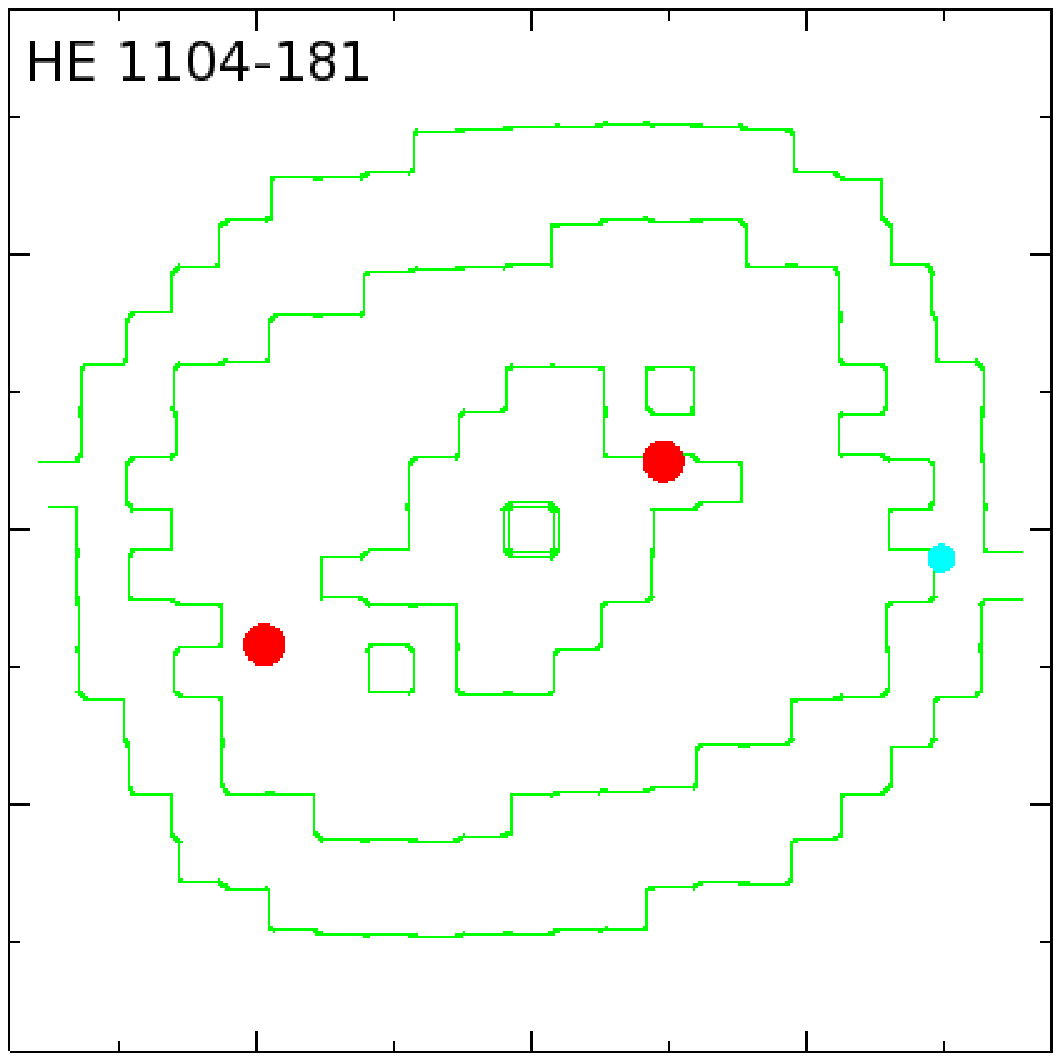}\\
\plotone{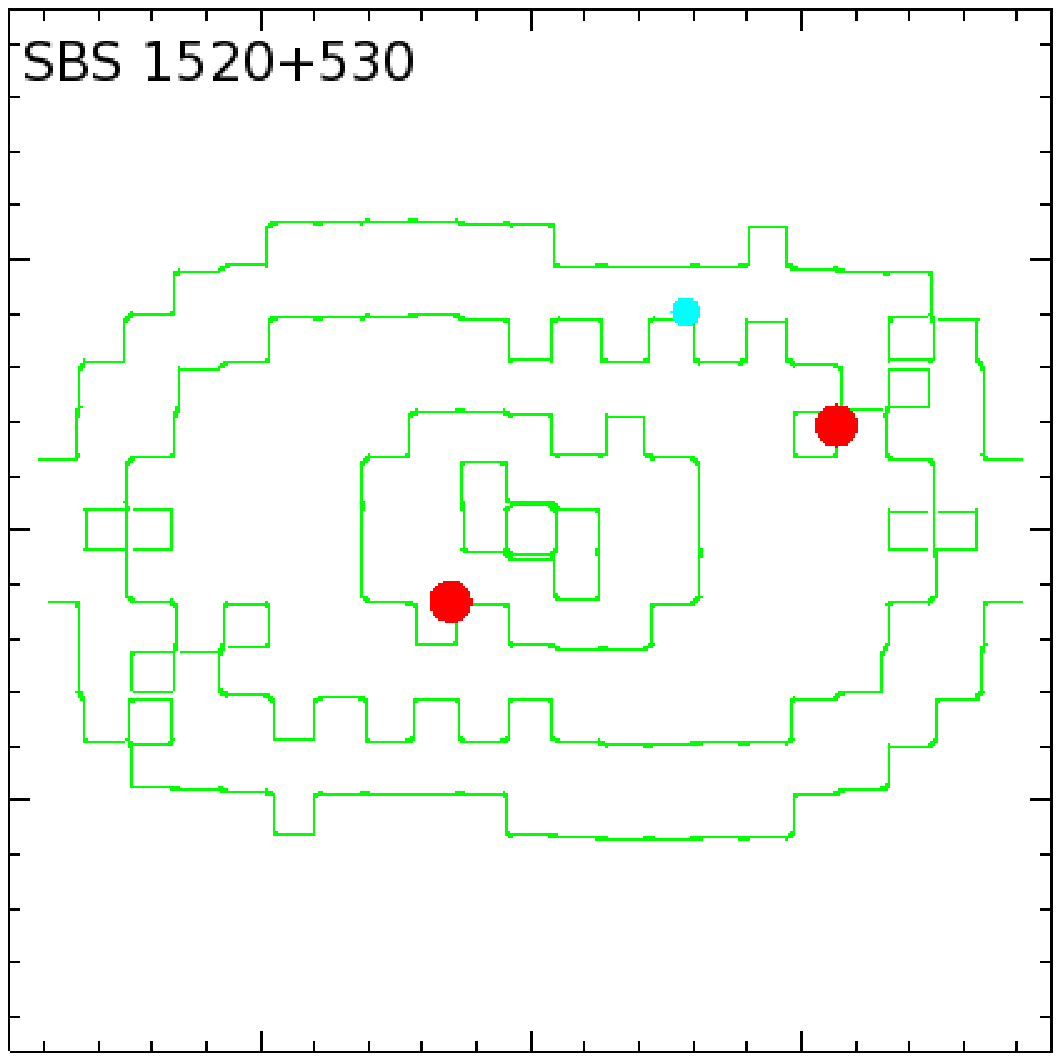}\plotone{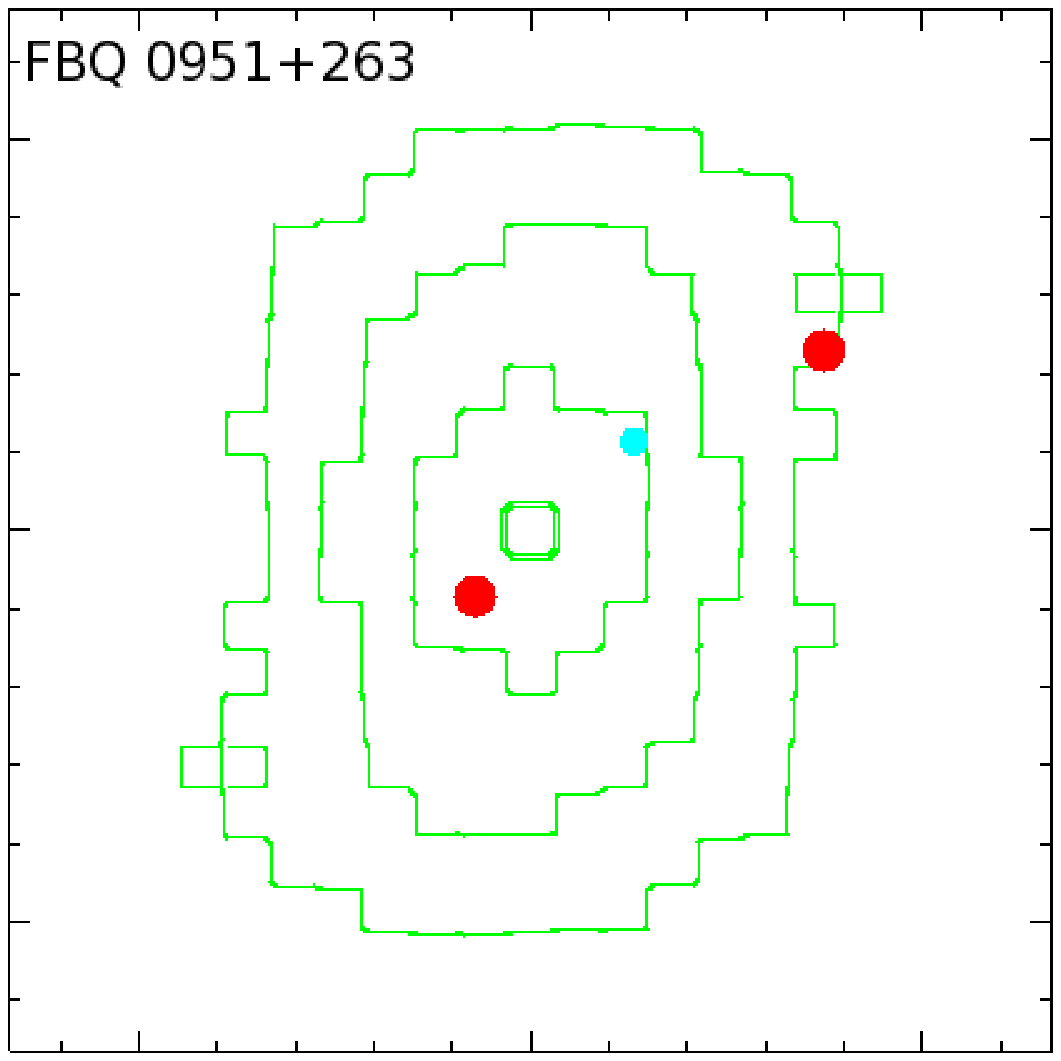}\plotone{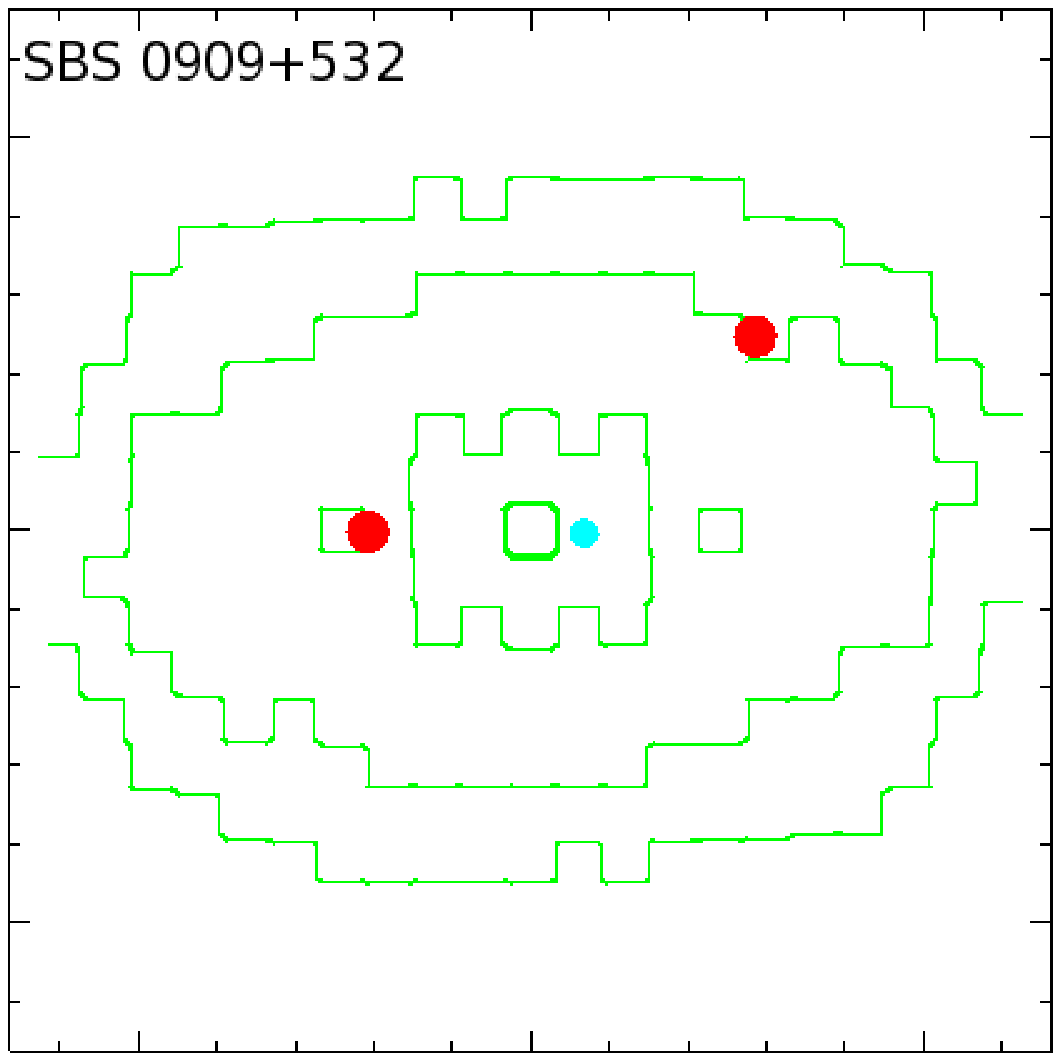}\\
\caption{Ensemble of  18 average mass maps of the lenses. The
larger tick marks in each panel correspond to $1\farcs0$. Red and cyan dots represent the positions of the images and the sources respectively. The contours are in logarithmic steps, with critical density corresponding to the third contour from the outside. The systems are grouped according their morphology described in the Section 5.}
\end{figure*}
\begin{deluxetable}{lllccrrrcl}
\tablewidth{0pt}
\tablecaption{Properties of time-delay lenses
}
\tablehead{
\colhead{System}  &
\colhead{$z_l$}&\colhead{$z_s$}&\colhead{P.A.}&\colhead{Point mass$^a$}&\colhead{x$^b$}&
\colhead{y$^b$}&\colhead{$\Delta\tau^c$}& \colhead{Reference}\\
\colhead{}&\colhead{}&\colhead{}&\colhead{}&\colhead{}&($\arcsec$)& ($\arcsec$)&(days)&
}
\startdata
\hline
PG 1115+080&0.311& 1.722 &$-45^\circ$&0& 0.381& 1.344&&1, 2, 3, 4\\
&&&&&$-$0.947& $-$0.69 &$13.3^{+0.9} _{-1.0}$\\ 
&&&&&$-$1.096 &$-$0.232 &0\\
&&&&&0.722& $-$0.617& $11.7^{+1.5} _{-1.6}$\\
 \hline
HE 2149-275 & 0.495 & 2.030&-&0&0.714& $-$1.150&&5, 6, 7\\
&&&&&$-$0.176 &0.296 &$103\pm12$\\
\hline
SDSS J1206+4332  &0.748 & 1.789 & $50^\circ$&0&0.663 & $-$1.749&&8, 9\\
&&&&&0.566& 1.146 & $116^{+4} _{-5}$\\
\hline
HE 0435-1223 & 0.454 &1.689& $-65^\circ$ &0&$-$1.165& 0.573&&10, 11\\
&&&&&1.311& $-$0.03& $2.1^{+0.78} _{-0.71}$\\ 
&&&&&0.311 &1.126& $6^{+1.07} _{-1.08}$\\
&&&&&$-$0.226& $-$1.041& $8.37^{+1.31} _{-1.37}$\\
\hline
SDSS J1650+4251 & 0.577 & 1.547 & $80^\circ$&0&0.017& 0.872&& 12, 13\\ 
&&&&&$-$0.206& $-$0.291&  $49.5\pm1.9$\\
 \hline
QSO J0957+561 & 0.356& 1.410&  $-30^\circ$ &0&  1.408&  5.034&& 14, 15, 16  \\
&&&&& 0.182 &$-$1.018  &$422.6\pm0.6$ \\
&&&&& 2.860  &3.470  &0\\
&&&&&$-$1.540 & $-$0.050 &   0\\
\hline
RX J1131-1231&0.295 &0.658&$-35^\circ$&0&$-$1.984 &0.578&& 17, 3\\
 &&&&&$-$1.426 &$-$1.73& $ 1.5 ^{+2.49} _{-2.02}$\\
&&&&&$-$2.016 &$-$0.610 & $9.61^{+1.97} _{-1.57}$\\
&&&&&1.096 &0.274 &$87\pm8$\\
\hline
RX J0911+055 &0.769&2.800&$90^\circ$ &1& 2.226&  0.278&& 18, 19, 20, 21\\ 
&&&&&$-$0.968 &$-$0.105 & $146\pm8$\\
&&&&&$-$0.709& $-$0.507  &  0 \\
&&&&&$-$0.696&  0.439  &  0\\
\hline
SDSS J1004+4112& 0.680& 1.734& $90^\circ$ &0&3.943 &$-$8.868 &&5, 22, 23, 24\\
&&&&&$-$8.412 &$-$0.86& $821.6\pm2.1$\\
&&&&&$-$7.095& $-$4.389& $40.6\pm1.8$\\ 
&&&&&1.303 &5.327 &0\\
 \hline
WFI J2033-4723 &  0.661 &1.660 & $0^\circ$&0&$-$1.439& $-$0.311&&25, 26, 27\\
&&&&&0.756& 0.949 &$35.5\pm1.4$\\ 
&&&&&0.044 &1.068 &0\\
&&&&&0.674& $-$0.5891& $27.1^{+ 4.1}_{-2.3}$\\ 
 \hline
PKS 1830-211 & 0.885 & 2.507 &-&0& $-$0.498& 0.456&& 28, 29, 30, 31\\
&&&&&0.151 &$-$0.268& $26^{+4} _{-5}$\\
 \hline
 B1600+434 &0.410 & 1.590  &$90^\circ$&1&0.610 & 0.814 && 5, 32, 33, 34\\
&&&&&$-$0.110 &$-$0.369& $51\pm4$\\
\hline
B0218+357 &0.685&0.944&- &0&0.250&$-$0.119 & & 35, 36, 37, 38\\
&&&&&$-$0.052&  0.007 &$10.1^{+1.5} _{-1.6}$& \\
B1608+656 & 0.630 & 1.394 &- & 1& $-$1.155& $-$0.896&& 39, 40, 41, 42\\
&&&&&$-$0.419 &1.059 & $31.5\pm1.5$ \\
&&&&&$-$1.165 & 0.610 & 0\\
&&&&&0.714 & $-$0.197 & $41\pm1.5$\\
\hline
HE 1104-181& 0.729& 2.319&-&2 &$-$1.936 &$-$0.832&& 43, 44, 45 \\
&&&&&0.965& 0.5& $152.2 ^{+2.8} _{-3.0}$\\
\hline
SBS 1520+530 &0.761&1.855 &-&1& 1.130 &  0.387&& 46, 47, 48, 49\\ 
&&&&&$-$0.296 & $-$0.265 &$130\pm3$\\
\hline
FBQ J0951+263 & 0.24&1.246&-&0& 0.750&  0.459 &&50, 51\\
&&&&&$-$0.142& $-$0.169&$16\pm 2$\\
\hline
SBS 0909+532 &0.830&1.376&-&0&0.572 &0.494&&5, 52, 53\\ 
&&&&&$-$0.415&$-$0.004&$45^{+1} _{-11}$\\
\hline
B1422+231& 0.339 &3.62 & $10^\circ$&0&1.014 &$-$0.168 && 54, 55, 3\\
&&&&&0.291 &0.900  &$7.6\pm2.5$\\
&&&&&0.680 &0.580 &$1.5\pm1.4$\\
&&&&&$-$0.271 &$-$0.222 &0\\
\enddata
\tablenotetext{a}{We have included all the galaxies that might have contribution in the lensing, we add them whenever one or more galaxies are visible in the field and when their redshift is similar to the main lens.}
\tablenotetext{b}{The positions of the QSO images are calculated with respect to the position of the main galaxy G1. The images are in arrival-time order. }
\tablenotetext{c}{All measured time delays are listed expect those for which error bars are large and, therefore, the detections are marginal.}
\tablerefs{
(1) \citet{Morgan:2008}; (2) \citet{Barkana:1997}; (3) \citet{Tonry:1998};  (4) \citet{Weymann:1980};
(5) CASTLES; (6) \citet{Burud:2002a}; (7) \citet{Wisotzki:1996};
(8) \citet{Paraficz:2009}; (9) \citet{Oguri:2005}; 
(10) \citet{Kochanek:2006}; (11) \citet{Morgan:2005};
(12) \citet{Morgan:2003}; (13) \citet{Vuissoz:2007}; 
(14)  \citet{Bernstein:1999}; (15)  \citet{Oscoz:2001}; (16) \citet{Falco:1997};
(17) \citet{Claeskens:2006}; 
(18) \citet{Burud:1998}; (19)  \citet{Hjorth:2002}; (20) \citet{Kneib:2000}; (21)  \citet{Bade:1997};
(22) \citet{Fohlmeister:2008}; (23) \citet{Inada:2003}; (24)  \citet{Williams:2004};
 (25) \citet{Vuissoz:2008}; (26) \citet{Eigenbrod:2006}; (27) \citet{Morgan:2004};
(28) \citet{Meylan:2005}; (29) \citet{Lovell:1998}; (30) \citet{Chengalur:1999}; (31) \citet{Lidman:1999}; 
(32) \citet{Burud:2000}; (33) \citet{Jaunsen:1997}; (34) \citet{Jackson:1995};
(35) \citet{Wucknitz:2004}; (36) \citet{Cohen:2000}; (37) \citet{Browne:1993}; \citet{Carilli:1993}; (38)  \citet{Cohen:2003}; 
(39) \citet{Koopmans:2003}; (40) \citet{Fassnacht:2002}; (41) \citet{Myers:1995}; (42) \citet{Fassnacht:1996};
(43) \citet{Poindexter:2007}; (44) \citet{Lidman:2000}; (45) \citet{Wisotzki:1993};
(46) \citet{Faure:2002}; (47) \citet{Burud:2002}; (48) \citet{Auger:2008}; (49) \citet{Chavushyan:1997}; 
(50) \citet{Jakobsson:2005}; (51)  \citet{White:2000}; 
(52) \citet{Ullan:2005}; (53) \citet{Lubin:2000}; 
(54) \citet{Raychaudhury:2003};  (55) \citet{Patnaik:2001};
}
\end{deluxetable}
\noindent
\citep{Oguri:2007} and 10 lenses \citep{Saha:2006}.  The $H_0$ 
distribution of the samples of 10 and 15 lenses are  slightly different than the original results of 
 \citet{Oguri:2007} and \citet{Saha:2006}\footnote{\citet{Oguri:2007} obtained $H_0=70\pm6$ $\rm{km}$ $\rm{s}^{-1} \rm{Mpc}^{-1}$, \citet{Saha:2006} got $H_0=72_{-11}^{+7}$ $\rm{km}$ $\rm{s}^{-1} \rm{Mpc}^{-1}$ }.
\citet{Oguri:2007}  used 16 lensed quasar systems (40 image pairs)
to constrain the Hubble constant. For each image pair, he computed the likelihood as a function of the Hubble constant. He then computed the effective $\chi^2$ by summing up the logarithm of the likelihoods. The first summation runs over lens systems, whereas the second summation runs over image pairs for each lens system. He derived the best-fit value and its error of Hubble constant in the standard way using a goodness-of-fit parameter.

\begin{figure}[!h]
\epsscale{1.0}
\plotone{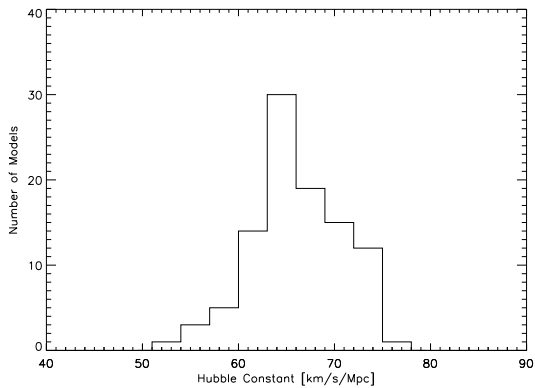}
\caption{Histogram of the ensembles of $H_0$ values estimated from 18 lensing systems. We have assumed $\Omega_\Lambda=0.7$ and $\Omega_m=0.3$\label{fig3}.}
\end{figure}
The difference with  \citet{Oguri:2007} is mainly due to use of different modeling method and a different statistical and modeling approach to obtain $H_0$. In the case of \citet{Saha:2006} the method  is identical and the difference  comes from
the use of other/newer data,  and the  use of different rules  for constraining shear, adding secondary lenses, etc. 
 Using the \citet{Saha:2006} sample of 10 lenses we have obtained $H_0= 63_{-5}^{+6}$ $\rm{km}$ $\rm{s}^{-1} \rm{Mpc}^{-1}$
and using the \citet{Oguri:2007} sample of 16 lenses (minus  B1422+231), we got $H_0=66_{-7}^{+5}$ $\rm{km}$ $\rm{s}^{-1} \rm{Mpc}^{-1}$.

\begin{figure}[!h]
\epsscale{1.0}
\plotone{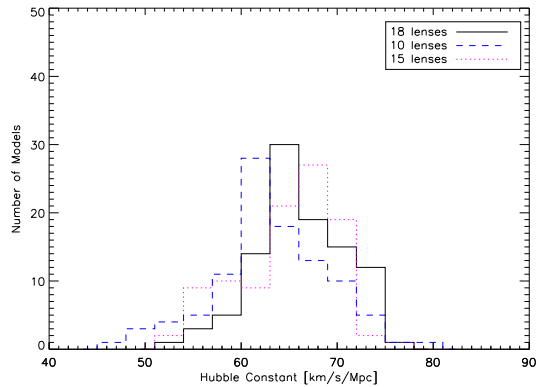}
\caption{Comparison of the three histograms of the ensembles of $H_0$ values. Histogram plotted with the solid line represents our results for 18 systems; the dashed line represents the results of 10 lenses \citep{Saha:2006} and the dotted line is a result of 16 lenses \citep{Oguri:2007}. All calculations are done for a flat Universe with $\Omega_\Lambda=0.7$ and $\Omega_m=0.3$. }
\end{figure}


\section{'Elliptical' sample}

The strongest degeneracy in lens modeling is the so-called mass-sheat degeneracy between time delays and the steepness of the mass profile. This degeneracy causes that without change in the images position, we can change the steepness of the mass profile, hence, the resulting $H_0$.
Thus, if the steepness and $\Delta \tau$ are known, $H_0$ is  well constrained, although the time delays are also influenced by other more complicated degeneracies  involving details of the shape of the lens  but these effects are secondary \citep{Saha:2006a}.

In our sample of 18 systems we have galaxies with a variety of steepness. 
Without detailed observations the profile steepness  of each lens is not known, except possibly for elliptical galaxies.

Several studies have shown that  elliptical galaxies  may be considered as approximately isothermal $\rho\propto r^{-2}$ \citep{Koopmans:2006,Oguri:2007,Koopmans:2009,Gerhard:2001}.
Hence, by selecting from our 18 system only elliptical galaxies we expect to get a uniform sample with known  slopes of mass profiles.

We have selected  five systems that are relatively isolated, elliptical galaxies: 
PG 1115+080 \citep{Impey:1998}, HE 2149-275 \citep{Lopez:1998}, SDSS J1206+4332  \citep{Paraficz:2009},
HE 0435-1223  \citep{Kochanek:2006} and SDSS J1650+4251 \citep{Vuissoz:2007,Morgan:2003}.

 Five other lenses were also found to be elliptical but a cluster or group to which they belong  has strong influence on the lensing system, thus it is difficult to model them: QSO 0957+561  \citep{Oscoz:1997,Bernstein:1999}; RX J1131-1231 \citep{Claeskens:2006}; RX J0911+055 \citep{Kneib:2000},  SDSS J1004+4112 \citep{Sharon:2005} and WFI J2033-4723  \citep{Eigenbrod:2006}, 

 Three of the lenses are most probably spiral galaxies: PKS 1830-211 is a face-on spiral galaxy \citep{Winn:2002}, B1600+434 is a spiral galaxy with a companion   \citep{Jaunsen:1997} and B0218+357 is an isolated spiral galaxy \citep{Koopmans:2001}. 
 
 The remaining  five systems were not included into the 'elliptical' sample due to various other issues:
 B1608+656 has two lensing galaxies inside the Einstein ring; HE 1104+180 has little starlight, suggesting a dark matter dominated lens \citep{Poindexter:2007,Vuissoz:2008}; SBS 1520+530  has a steeper than isothermal slope, probably due to mergers \citep{Auger:2008};
FBQ 0951+263 is  a complicated system which is hard to model \citep{Peng:2006};
SBS 0909+532  is probably early-type \citep{Lehar:2000}  but according  \citet{Motta:2002} is not very typical due to lots of dust.

Using the  sample of five elliptical galaxies and constraining the steepness of their mass profiles to be $\alpha_{min}=1.8$ and $\alpha_{max}=2.2$ we have run the PixeLens simultaneous modeling.
The sample of 5 lensing systems gives us a Hubble constant estimation $H_0=79_{ -3}^{+2}$ $\rm{km}$ $\rm{s}^{-1} \rm{Mpc}^{-1}$ at 68\% confidence and $H_0= 79_{ -4}^{+3}$ $\rm{km}$ $\rm{s}^{-1} \rm{Mpc}^{-1}$ at 90\% confidence.
We have also combined the 5 constrained systems with the  rest of the unconstrained sample to perform simultaneous modeling  and obtained very well determined Hubble constant
$H_0=76\pm3$ $\rm{km}$ $\rm{s}^{-1} \rm{Mpc}^{-1}$  at 68\% confidence and  $H_0=76\pm5$ $\rm{km}$ $\rm{s}^{-1} \rm{Mpc}^{-1}$  at 90\% confidence.
The results are presented in Figure 4.

\begin{figure}[!h]
\epsscale{.80}
\plotone{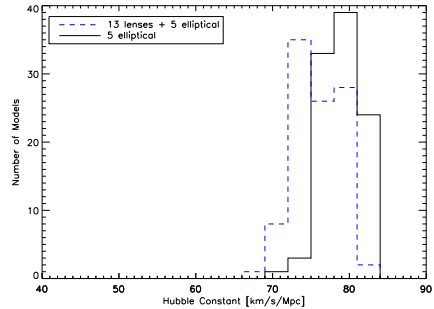}
\caption{Histogram of the distribution of $H_0$ values for  the selected sample of 5 elliptical galaxies having constrained steepness of mass profiles in the range 1.8--2.2 (solid line) and the selected elliptical sample combined  with the rest of the systems (dashed line). All calculations are done for a flat Universe with $\Omega_\Lambda=0.7$ and $\Omega_m=0.3$. }
\end{figure}


\section{Conclusions}


Non-parametric modeling using PixeLens was applied to an ensemble of 18 lenses to determine a new value for the Hubble constant. We have obtained $H_0=66_{-4}^{+6}$ $\rm{km}$ $\rm{s}^{-1} \rm{Mpc}^{-1}$ for a flat Universe with $\Omega_\Lambda=0.7$, $\Omega_m=0.3$. 
We have also compared our results with the two previous  attempts to estimate $H_0$ from time delays \citep{Saha:2006,Oguri:2007} (Fig. 3).

Our additional result was based on studies of a selected sample of five lensing galaxies that have mass profiles close to $\rho\propto r^{-2}$. 
The Hubble constant recovered using the selected sample of elliptical galaxies combined  with the  rest of the systems is $H_0=76_{ -3}^{+3}$  $\rm{km}$ $\rm{s}^{-1} \rm{Mpc}^{-1}$. 

The gravitational lensing method that constrains $H_0$ has difficulties due to a couple of degeneracies between mass and time delay. The major degeneracy, the mass-sheet degeneracy, as we have shown,  can be addressed by a careful choice  of galaxies and the others partially  by a combined, pixelated analysis of a large sample of lenses. Pixelated lens modeling  provides insight into the structure of galaxies and the distribution of dark matter which together with precise measurements of time delays gives a reliable cosmological method. 

Lensing can already determine the Hubble constant approaching  the accuracy level of other leading measurements. Nevertheless, still more observation are needed.
Future data  with precise time delay measurements and better lens models will give  even better constraints on $H_0$,  perhaps turning lensing into a very competitive method.

\acknowledgments
The Dark Cosmology Centre is funded by the DNRF. DP thanks  Prasenjit Saha for very extensive and patient help with PixeLens.

\bibliographystyle{apj}
\bibliography{paraficz_danka}

\end{document}